\documentclass[11pt]{article}
\usepackage{amssymb}

\catcode`@=11
\def\secteqno{\@addtoreset{equation}{section}%
\def\theequation{\thesection.\arabic{equation}}}
\catcode`@=12
\secteqno

\begin{document}
\setlength{\baselineskip}{5.5mm}

\thispagestyle{empty}

\begin{flushright}
 KEK-TH-933
\end{flushright}

\vspace*{10mm}

\begin{center}
 {\Large \bf Note on Gauge Theory on Fuzzy Supersphere}

\vspace{10mm}
{\large Satoshi Iso and Hiroshi Umetsu}

\vspace{5mm}

{\it 
Theory Division,\ High Energy Accelerator Research Organization (KEK),\\
Tsukuba,\ Ibaraki,\ 305-0801, Japan }

\vspace{5mm}

{\small\sf E-mails:\ satoshi.iso@kek.jp, umetsu@post.kek.jp} 
\end{center}

\vspace{10mm}

\begin{abstract}
We construct a supermatrix model whose classical background 
gives two-dimensional noncommutative supersphere.
Quantum fluctuations around it give the supersymmetric 
gauge theories on the fuzzy supersphere 
constructed by Klimcik.
This model has a parameter $\beta$ which can tune masses of the 
particles in the model and interpolate various supersymmetric 
gauge theories on sphere.

\end{abstract}

\newpage

\setcounter{page}{1}
\parskip 2mm

\section{Introduction}
Matrix models have been 
vigorously studied to understand nonperturbative
aspects of string theories.
In matrix models of the IKKT type~\cite{IKKT}, background space-time
appears dynamically as a classical background of matrices and 
their fluctuations around the classical solution are regarded as gauge
and matter fields on the space-time.
In particular,  matrix models describe noncommutative gauge theories
when the classical background of matrices are noncommutative
\cite{AIIKKT}. 
In this approach constructions of the open Wilson loop and background
independence of the noncommutative gauge theories are 
clarified \cite{IIKK}.

Noncommutative gauge theories appear on D-branes in string theories
in a constant NS-NS two form $B$ background
\cite{Schomerus, SeibergWitten} where the bosonic space-time coordinates
become noncommutative.  
Recently it was suggested that the non anti-commutativity of the
fermionic coordinates on the superspace appears in string theories in
a background of the $RR$ or graviphoton field strength
\cite{OV, deBoer, Seiberg}.
Since the non anti-commutative fermionic coordinates can be described by 
(super)matrices as well the noncommutative bosonic coordinates, it is
expected that (super)matrix models play an important role to investigate 
various aspects of field theories on the noncommutative superspace.
There are analyses of noncommutative superspace by using
supermatrices~\cite{GKP}-\cite{IU}.  
Supersymmetric actions for scalar multiplets on the fuzzy
two-supersphere were constructed in \cite{GKP} based on the 
$osp(1|2)$ graded Lie algebra. Furthermore a graded differential
calculus on the fuzzy supersphere is discussed in \cite{Grosse2}.
Supersymmetric gauge theories on this noncommutative superspace was
studied in \cite{Klimcik} by using differential forms on it.
In \cite{HIU}, noncommutative superspaces and their flat limits 
are studied by using the graded Lie algebras $osp(1|2)$, $osp(2|2)$ 
and $psu(2|2)$. 
Recently the concept of noncommutative superspace based on a
supermatrix was also introduced in proving the Dijkgraaf-Vafa
conjecture as the large N reduction~\cite{Kawai}.
Supermatrix model was also studied from the viewpoint of background 
independent formulations of matrix model which are expected to
give constructive definitions of string theories~\cite{supermatrix}.

In the previous paper~\cite{IU}, we constructed a supersymmetric gauge
theory on a fuzzy two-supersphere based on a supermatrix model.
This model has a classical solution representing a fuzzy supersphere
and we obtained a supersymmetric gauge theory on the fuzzy
two-supersphere expanding the supermatrices around the classical
background.
In a commutative limit this model  is , however,
different from the ordinary 
gauge theory in $D=2$,  e.g., the action includes higher derivative
terms and the fermions transform as spin $\frac{3}{2}$ 
representation under the $su(2)$ isometry group on $S^2$.
These differences are originated from the fact that we did not impose 
appropriate constraints on the supermatrices to eliminate extra degrees
of freedom. 
Supermatrix formulation of supersymmetric gauge theories 
is similar to the covariant superspace approach 
for the ordinary supersymmetric gauge theories~\cite{superspace}
because each supermatrix corresponds to the connection superfields
on the superspace as we saw in our previous paper \cite{IU}.
In the covariant superspace approach 
various constrains are imposed on the connections
to eliminate redundant degrees of freedom, 
but in our model 
we could not impose appropriate conditions.
After we wrote the previous paper \cite{IU}, 
we were noticed the
Klimcik's paper~\cite{Klimcik} where
he constructed a  supersymmetric gauge theory on the fuzzy
supersphere by using a method of differential forms 
and imposing suitable constraints on the connection superfields.
A crucial point of his construction 
is  the use of the enlarged $osp(2|2)$  algebra.
The global ${\cal N}=1$ supersymmetry algebra 
on the fuzzy supersphere is $osp(1|2)$. 
By adding the covariant derivatives 
$osp(1|2)$ is enlarged to $osp(2|2)$ 
because the supersymmetry generators and the covariant derivatives do
not anti-commute on the fuzzy supersphere. 
$Osp(2|2)$ algebra can be regarded as ${\cal N}=2$ superalgebras on the
sphere. 
Starting from the connection superfields on this ${\cal N}=2$
superspace, he found the correct constraints 
to obtain a supersymmetric gauge theory on the fuzzy supersphere.

In this paper we reformulate Klimcik's gauge theory on fuzzy
supersphere in terms of supermatrix models.
Very interestingly, Klimcik's gauge theory can be obtained from a
supermatrix model whose classical solution gives noncommutative
supersphere. 
Its quantum fluctuations become the supersymmetric gauge theory
proposed by him.
In a commutative limit, by taking a Wess-Zumino like gauge,
this model becomes the ordinary 
supersymmetric gauge theory on $S^2$. 
In the paper we use the manifestly $SO(3)$ covariant 
coordinates and decompose the bosonic field $a_i$ into
the normal component $\phi$ and tangential component $a_i^{(2)}$
on the sphere. The action contains a parameter which can 
tune masses of particles in the model.

\section{$\mathbf{osp(1|2)}$ and $\mathbf{osp(2|2)}$ algebras}

The graded commutation relations of the $osp(2|2)$ algebra are given by 
\begin{eqnarray}
 \begin{array}{lll}
  \left[\hat{l}_i, \hat{l}_j\right]=i\epsilon_{ijk}\hat{l}_k, &
   \left[\hat{l}_i, \hat{v}_\alpha\right]
   =\frac{1}{2}\left(\sigma_i\right)_{\beta\alpha}\hat{v}_\beta, &
   \left\{\hat{v}_\alpha, \hat{v}_\beta\right\}
   =\frac{1}{2}\left(C\sigma_i\right)_{\alpha\beta}\hat{l}_i, \\
   \left\{\hat{v}_\alpha, \hat{d}_\beta\right\}
    =-\frac{1}{4}C_{\alpha\beta}\hat{\gamma}, &
    \left[\hat{l}_i, \hat{d}_\alpha\right]
   =\frac{1}{2}\left(\sigma_i\right)_{\beta\alpha}\hat{d}_\beta, & 
   \left\{\hat{d}_\alpha, \hat{d}_\beta\right\}
   =-\frac{1}{2}\left(C\sigma_i\right)_{\alpha\beta}\hat{l}_i, \\
  \left[\hat{\gamma}, \hat{v}_\alpha\right]=\hat{d}_\alpha, &
   \left[\hat{\gamma}, \hat{d}_\alpha\right]=\hat{v}_\alpha, &
   \left[\hat{\gamma}, \hat{l}_i\right]=0,
 \end{array}
\end{eqnarray}
where $\hat{l}_i \ (i=1, 2, 3)$ and $\hat{\gamma}$ are bosonic
generators, and  $\hat{v}_\alpha$ and $\hat{d}_\alpha \ (\alpha =1, 2)$
are fermionic ones. $C=i\sigma_2$ is the charge conjugation matrix.
The $osp(1|2)$ subalgebra consists of the generators $\hat{l}_i$ and
$\hat{v}_\alpha$.
Irreducible representations of the $osp(1|2)$
algebra~\cite{representation}  
are characterized by values of the second Casimir operator 
$\hat{K}_2=\hat{l}_i\hat{l}_i
+C_{\alpha\beta}\hat{v}_\alpha \hat{v}_\beta =L(L+\frac{1}{2})$ 
where quantum number $L \in {\mathbb Z_{\geq 0}/2}$ is called super
spin. 
Each representation of $osp(1|2)$ consists of spin $L$ and
$L+\frac{1}{2}$ representations of the $su(2)$ subalgebra generated by
$\hat{l}_i$. The dimension of the representation is $N\equiv 4L+1$.
These representations are also the 
so-called 'atypical representations' of $osp(2|2)$ where 
$osp(1|2)$ algebra can be enlarged to $osp(2|2)$ algebra
by adding extra generators with the same size matrices.
The representation matrices of the generators $\hat{d}_\alpha$ and
$\hat{\gamma}$ can be written as second order polynomials of the
super spin $L$ representation matrices $l^{(L)}_i$ and $v^{(L)}_\alpha$
of the $osp(1|2)$ generators,
\begin{eqnarray}
 &&  \label{gamma}
  \gamma^{(L)}=-\frac{1}{L+1/4}
  \left(C_{\alpha\beta}v^{(L)}_\alpha v^{(L)}_\beta 
   +2L\left(L+\frac{1}{2}\right)\right),  \\
 && \label{d}
  d^{(L)}_\alpha = \frac{1}{2(L+1/4)}(\sigma_i)_{\beta\alpha}
  \left\{l^{(L)}_i, v^{(L)}_\beta\right\}.
\end{eqnarray}

The condition $\hat{K}_2=L(L+1/2)$ defines a two-dimensional
supersphere. Consider polynomials $\Phi(l^{(L)}_i, v^{(L)}_\alpha)$ of
the representation matrices of $l^{(L)}_i$ and $v^{(L)}_\alpha$ with
super spin $L$.
These polynomials are $(4L+1)\times(4L+1)$ supermatrices.
Let us denote the space spanned by $\Phi(l^{(L)}_i, v^{(L)}_\alpha)$ as
${\cal A}_L$.
The $osp(1|2)$ algebra acts on ${\cal A}_L$. 
In particular we denote the adjoint action of the $osp(1|2)$ generators
as 
\begin{equation}
 \hat{\cal L}_i\Phi = \left[l^{(L)}_i, \Phi\right], \qquad
  \hat{\cal V}_\alpha\Phi = \left[v^{(L)}_\alpha, \Phi\right].
\end{equation} 
${\cal A}_L$ can be decomposed into irreducible representations
under the adjoint action of the $osp(1|2)$ algebra as
$0\oplus \frac{1}{2}\oplus 1 \oplus\cdots\oplus 2L-\frac{1}{2}\oplus2L$. 
The dimension of this space is $(4L+1)^2$ and thus any
$(4L+1)\times(4L+1)$ matrices can be expanded by these polynomials.
Useful basis for the expansion are the matrix super spherical harmonics
$Y^S_{km}(l^{(L)}_i, v^{(L)}_\alpha)$, 
\begin{eqnarray}
 && \left(\hat{\cal L}_i\hat{\cal L}_i
     +C_{\alpha\beta}\hat{\cal V}_\alpha\hat{\cal V}_\beta\right)
 Y^S_{km}(l^{(L)}_i, v^{(L)}_\alpha)
 = k\left(k+\frac{1}{2}\right)Y^S_{km}(l^{(L)}_i, v^{(L)}_\alpha), \\
 && \hat{\cal L}_3 Y^S_{km}(l^{(L)}_i, v^{(L)}_\alpha)
  =mY^S_{km}(l^{(L)}_i, v^{(L)}_\alpha).
\end{eqnarray}  
Then $\Phi(l^{(L)}_i, v^{(L)}_\alpha)$ can be expressed as a series of
the super spherical harmonics,
\begin{equation}
 \Phi(l^{(L)}_i, v^{(L)}_\alpha)
  =\sum_{k=0, 1/2, 1, \cdots}^{2L}
  \phi_{km}Y^S_{km}(l^{(L)}_i, v^{(L)}_\alpha),
\end{equation}
where the Grassmann parity of the coefficient $\phi_{km}$ is determined
by the grading of the spherical harmonics.
We can map the supermatrix $\Phi(l^{(L)}_i, v^{(L)}_\alpha)$ to a function
on the superspace $(x_i, \theta_\alpha)$ by
\begin{equation}
 \Phi(l^{(L)}_i, v^{(L)}_\alpha) \longrightarrow 
  \phi(x_i, \theta_\alpha)
  =\sum_{k,m}\phi_{km}y^S_{km}(x_i, \theta_\alpha), 
\end{equation}
where $y^S_{km}(x_i, \theta_\alpha)$ is the ordinary super spherical
function. A product of supermatrices is mapped to a noncommutative star
product of functions on the fuzzy supersphere \cite{Balachandran}.

In constructions of field theories on the fuzzy supersphere, 
the $osp(2|2)$ generators $\hat{d}_\alpha$ and $\hat{\gamma}$ play 
important roles. 
The adjoint action of $\hat{d}_\alpha$ on supermatrices is the covariant
derivative on the fuzzy supersphere.
The kinetic terms for a scalar multiplet on the supersphere are
constructed by using these generators~\cite{GKP}.

\section{Supersymmetric gauge theory on fuzzy supersphere}
In this section we reformulate the Klimcik's 
construction of  a supersymmetric gauge theory on fuzzy supersphere 
\cite{Klimcik} in terms of supermatrices.
We construct a supermatrix model which has a classical
solution representing the fuzzy supersphere and
expanding supermatrices around this background we will obtain a
supersymmetric gauge theory given by Klimcik.
This formulation has similarities to the covariant superspace approach 
in the ordinary supersymmetric gauge theory. 
Namely, we first
introduce larger degrees of freedom 
corresponding to the connection 
superfields on the fuzzy supersphere.  
In order to eliminate redundant degrees of freedom,
we need to impose 
appropriate constraints on them.
After fixing extra gauges, we will
obtain necessary degrees of freedom to
describe a 
supersymmetric gauge theory on the noncommutative supersphere. 

We first consider
direct products of two vector spaces of supermatrices,
${\cal A}_L$ and ${\cal A}_{L'}$.
$L'$ can be taken as any superspin.
On the other hand, $L$ should be taken large in order to take 
a commutative limit. 
Each element  is a $(4L+1)(4L'+1) \times
(4L+1)(4L'+1)$ supermatrix. we then restrict to consider
a special type of supermatrices which can be written as
\begin{equation}
M = \sum_A X_{A} \otimes T^A
\end{equation}
where $X_A \in {\cal A}_L$ is a general supermatrix 
and $T^A$ is the superspin $L'$
representation matrix of the $osp(2|2)$ generators;
$T^A \ =\{l^{(L')}_i, v^{(L')}_\alpha, d^{(L')}_\alpha,$
$\gamma^{(L')}\}$. 
Among them, we can define two kinds of products, $\cdot$ and $\ast$, 
\begin{eqnarray}
 && \left(X_A \otimes T^A\right) \cdot \left(Y_B \otimes T^B\right)
  = (-1)^{|T^A||Y_B|} X_A Y_B \otimes T^A T^B, \\
 && \label{*}
  \left(X_A \otimes T^A\right) \ast \left(Y_B \otimes T^B\right)
  = (-1)^{|T^A||Y_B|} X_A Y_B \otimes \left[T^A, T^B\right\},
\end{eqnarray}
where $|T^A|$ and $|Y_B|$ are the gradings of $T^A$ and $Y_B$
respectively. 
\footnote{In the paper \cite{Klimcik}, $*$ is meant for 
$*^{G}-*^{H}$ but in our paper  we use 
$*$ product as  $*^{G}$ or $*{^H}$ 
according to operated supermatrices. }

The supermatrix $M$ defined above can be expanded explicitly as
\begin{eqnarray}
\label{M}
 M=A_i\otimes l^{(L')}_i 
  + C_{\alpha\beta}\varphi_\alpha \otimes v^{(L')}_\alpha
  - C_{\alpha\beta}\psi_\alpha \otimes d^{(L')}_\beta 
  - \frac{1}{4}W \otimes \gamma^{(L')}.
\end{eqnarray}
Here $A_i$ and $W$ are $(4L+1) \times (4L+1)$ even supermatrices, and
$\varphi_\alpha$ and $\psi_\alpha$ are $(4L+1) \times (4L+1)$ odd
supermatrices. 
$L'$ is arbitrary thus one may set $L' = \frac{1}{2}$ for simplicity.
$M$ satisfies a reality condition $M^\ddagger = M$, that is 
$A_i^\ddagger = A_i, 
\varphi_\alpha^\ddagger = C_{\alpha\beta}\varphi_\beta, 
\psi_\alpha^\ddagger = C_{\alpha\beta}\psi_\beta$ and
$W^\ddagger = W$. 
The definition of $\ddagger$ is referred to the appendix of \cite{IU}.
We denote the $osp(1|2)$ part of $M$ as $M_H$
\begin{equation}
 M_H=A_i\otimes l^{(L')}_i 
  + C_{\alpha\beta}\varphi_\alpha \otimes v^{(L')}_\alpha,
\end{equation}
and the rest as $M_{H^\perp} \equiv M-M_H$.
Since all the components $A_i, \phi_{\alpha}, \psi_{\alpha}$ and $W$
are supermatrices and can be expanded in terms of super spin $L$
spherical harmonics, they can be regarded as superfields on 
the fuzzy supersphere.
The supermatrix $M$ is shown to correspond to the covariant derivatives
on the ${\cal N}=2$ extended supersphere. 
Then we define 'field strength' as
\begin{eqnarray}
 \label{field strength}
 F &=& M \ast M - M_H \ast M_H + \frac{d_G}{2} M 
 - \frac{d_H}{2} M_H \nonumber \\
 &=& \left(\frac{1}{4}(\sigma_iC)_{\alpha\beta}
      \left\{\psi_\alpha, \psi_\beta\right\}
      -\frac{1}{4}A_i\right)\otimes l^{(L')}_i
 +\left(-\frac{1}{4}C_{\alpha\beta}\left[\psi_\alpha, W\right]
   -\frac{1}{4}C_{\alpha\beta}\varphi_\alpha\right)\otimes v^{(L')}_\beta
 \nonumber \\
 && +\left(\frac{1}{2}(\sigma_iC)_{\alpha\beta}\left[A_i, \psi_\alpha\right]
  +\frac{1}{4}C_{\alpha\beta}\left[\varphi_\alpha, W\right]
  -\frac{1}{2}C_{\alpha\beta}\psi_\alpha\right)\otimes d^{(L')}_\beta 
 \nonumber \\
 &&+\left(-\frac{1}{4}C_{\alpha\beta}
   \left\{\varphi_\alpha, \psi_\beta\right\}
   -\frac{1}{8}W\right)\otimes\gamma^{(L')},
\end{eqnarray}
where $d_G=1$ and $d_H=\frac{3}{2}$ are Dynkin numbers.

In order to eliminate redundant superfields,
we follow the prescription in \cite{Klimcik}
and impose a constraint for the $osp(1|2)$ part of the field
strength to vanish,
\begin{eqnarray}
 \label{constraint1}
 F|_H \equiv 
   \left(\frac{1}{4}(\sigma_iC)_{\alpha\beta}
      \left\{\psi_\alpha, \psi_\beta\right\}
      -\frac{1}{4}A_i\right)\otimes l^{(L')}_i
 +\left(-\frac{1}{4}C_{\alpha\beta}\left[\psi_\alpha, W\right]
   -\frac{1}{4}C_{\alpha\beta}\varphi_\alpha\right)\otimes v^{(L')}_\beta 
 = 0.
\end{eqnarray}
This constraint breaks the $osp(2|2)$ covariant structure, but still
preserves the covariance under $osp(1|2)$ of the model. 
From this constraint,
 $A_i$ and $\varphi_\alpha$ can be solved in terms of $\psi_\alpha$
and $W$ as
\begin{eqnarray}
 && A_i = (\sigma_iC)_{\alpha\beta}
      \left\{\psi_\alpha, \psi_\beta\right\}, \\
 && \varphi_\alpha = -\left[\psi_\alpha, W\right].
\end{eqnarray}
Moreover we need to constrain further redundant
 degrees of freedom $W$ by the following condition,
\begin{eqnarray}
 \label{constraint2}
 -\frac{1}{L'(L'+1/2)}{\rm STr}_{L'} \left(M_{H^\perp}\cdot M_{H^\perp}\right)
 =C_{\alpha\beta}\psi_\alpha\psi_\beta + \frac{1}{4}W^2
 =L(L+1/2),
\end{eqnarray}
where ${\rm STr}_{L'}$ means taking a supertrace with respect to the
super spin $L'$ representation matrices.
This constraint is also $osp(1|2)$ invariant. 
In a commutative limit, this equation can be solved 
to eliminate $W$, as  will be seen. 

 We now start from the following action for the supermatrix $M$; 
\begin{eqnarray}
 && S=S_{F^2} + \beta S_{CS}, \\
 && S_{F^2} = {\rm STr}\left(F \cdot F\right), \\
 && S_{CS} = {\rm STr} \left(\frac{2}{3}M \cdot (M \ast M)
			-\frac{2}{3}M_H\cdot (M_H \ast M_H) 
		     +\frac{d_G}{2} M \cdot M 
		     - \frac{d_H}{2}M_H \cdot M_H\right), 
\end{eqnarray}
where $\beta$ is a real constant parameter and $d_G$ and 
$d_H$ were defined above. 
This action is invariant under the following $osp(1|2)$ 
supersymmetry transformation, 
\begin{equation}
 \label{osp12}
 \delta M=\left[C_{\alpha\beta}\lambda_\beta\otimes v^{(L')}_\alpha, \
	   M\right], 
\end{equation}
where a parameter $\lambda_\alpha$ is a $(4L+1)\times (4L+1)$ grading
matrix multiplied by a Grassmann parameter $\tilde{\lambda}_\alpha$,
\begin{equation}
 \lambda_\alpha = \tilde{\lambda}_\alpha
  \left(
   \begin{array}{cc}
    1_{2L+1} & 0 \\
    0 & -1_{2L}
   \end{array}
  \right).
\end{equation}
In addition to the supersymmetry, this action has invariance under
the $u(2L+1|2L)$ gauge transformation,
\begin{equation}
 \delta M = \left[M, \ H\otimes 1\right],
\end{equation}
where $H\in u(2L+1|2L)$.
We note that $S_{F^2}$ and $S_{CS}$ are independently invariant under
the supersymmetry and gauge transformations.

Then we solve the equation of motion of the above action 
to obtain a classical solution and expand supermatrices $M$
around it.  The classical solution represents the 
background space-time and fluctuations are regarded as 
gauge fields on the space-time.
One of the classical solutions is given by
\begin{equation}
 \label{background}
 A_i=l^{(L)}_i, \quad \varphi_\alpha = v^{(L)}_\alpha, \quad
  \psi_\alpha = d^{(L)}_\alpha, \quad W = \gamma^{(L)},
\end{equation}
which represents a fuzzy supersphere.
It should be noted that the solution satisfies not only the equations of
motion but also the constraints (\ref{constraint1}) and (\ref{constraint2}). 
$M$ can be decomposed into the classical background (\ref{background})
and fluctuation $\tilde{M}$ as
\begin{eqnarray}
 M &=& M_{cl} + \tilde{M}, \\
 M_{cl} &=& l^{(L)}_i\otimes l^{(L')}_i 
  + C_{\alpha\beta}v^{(L)}_\alpha\otimes v^{(L')}_\beta
  - C_{\alpha\beta}d^{(L)}_\alpha\otimes d^{(L')}_\beta
  - \frac{1}{4}\gamma^{(L)}\otimes \gamma^{(L')}.
\end{eqnarray}
Because of the $osp(2|2)$ algebra
the field strength for the classical background
$M_{cl}$ vanishes,
\begin{eqnarray}
F_{cl}= M_{cl}\ast M_{cl} - M_{H cl}\ast M_{H cl} 
  +\frac{d_G}{2} M_{cl} - \frac{d_H}{2} M_{H cl}=0,
\end{eqnarray}
where $M_{H cl}$ is the $osp(1|2)$ part of $M_{cl}$.
Then the field strength (\ref{field strength}) becomes
\begin{eqnarray}
 F &=& 
  \left(M_{cl}\ast \tilde{M} + \tilde{M}\ast M_{cl}
   +\frac{d_G}{2}\tilde{M}\right)
  -\left(M_{H cl}\ast \tilde{M}_H + \tilde{M}_H\ast M_{H cl}
    +\frac{d_H}{2}\tilde{M}_H\right) \nonumber \\
 && + \tilde{M}\ast\tilde{M}-\tilde{M}_H\ast\tilde{M}_H,
\end{eqnarray}
where $\tilde{M}_H$ is the $osp(1|2)$ part of $\tilde{M}$.
In this form, terms in the first and second parenthesises 
respectively coincide with $\delta^{G} \tilde{M}$ and 
$\delta^{H} \tilde{M}_H$ in Klimcik's notation \cite{Klimcik} 
and thus
this is nothing but the field strength defined by him.  

Expanding each component of the supermatrices around the classical
background,  
\begin{eqnarray}
 A_i = l^{(L)}_i+\tilde{A}_i, \quad 
  \varphi_\alpha = v^{(L)}_\alpha+\tilde{\varphi}_\alpha, \quad
  \psi_\alpha = d^{(L)}_\alpha+\tilde{\psi}_\alpha, \quad
  W=\gamma^{(L)}+\tilde{W},
\end{eqnarray}
$F|_{H^\perp}$ and $S_{CS}$ become
\begin{eqnarray}
 F|_{H^\perp}&=&
  \left[\frac{1}{2}(\sigma_i C)_{\alpha\beta}
  \left([l^{(L)}_i, \tilde{\psi}_\alpha]-[d^{(L)}_\alpha, \tilde{A}_i]
  \right)
  +\frac{1}{4}C_{\alpha\beta}
  \left(
   [v^{(L)}_\alpha, \tilde{W}]-[\gamma^{(L)}, \tilde{\varphi}_\alpha] 
  \right)
  -\frac{1}{2}C_{\alpha\beta}\tilde{\psi}_\alpha\right]
  \otimes d^{(L')}_\beta \nonumber \\
 && +\left[
      -\frac{1}{4}C_{\alpha\beta}
      \left(\{v^{(L)}_\alpha, \tilde{\psi}_\beta\}
      +\{d^{(L)}_\beta, \tilde{\varphi}_\beta\}\right)
      -\frac{1}{8}\tilde{W}
     \right]\otimes\gamma^{(L')},
 \\ 
 S_{CS} &=& \frac{1}{4}L'(L'+1/2){\rm STr}_L
  \left(
   2(\sigma_i C)_{\alpha\beta}\tilde{A}_i
   \{\tilde{\psi}_\alpha, \tilde{\psi}_\beta\}
   -2C_{\alpha\beta}\tilde{W}
   \{\tilde{\varphi}_\alpha, \tilde{\psi}_\beta\}
   \right. \nonumber \\
  && \left. -2(\sigma_i C)_{\alpha\beta}\tilde{\psi}_\alpha
  [l^{(L)}_i, \tilde{\psi}_\beta]
  +4(\sigma_i C)_{\alpha\beta}\tilde{A}_i
  \{d^{(L)}_\alpha, \tilde{\psi}_\beta\} \right. \nonumber \\
  && \left. +2C_{\alpha\beta}\tilde{\varphi}_\alpha
  [\gamma^{(L)}, \tilde{\psi}_\beta]
  -2C_{\alpha\beta}\tilde{W}\{v^{(L)}_\alpha, \tilde{\psi}_\beta\}
  -2C_{\alpha\beta}\tilde{W}\{d^{(L)}_\beta, \tilde{\varphi}_\alpha\}
  \right. \nonumber \\
 && \left. -\tilde{A}_i\tilde{A}_i
  -C_{\alpha\beta}\tilde{\varphi}_\alpha\tilde{\varphi}_\beta
  -2C_{\alpha\beta}\tilde{\psi}_\alpha\tilde{\psi}_\beta
  -\frac{1}{2}\tilde{W}^2
  \right).
\end{eqnarray}
Note that the supermatrices $A_i, \phi_{\alpha}, \psi_{\alpha}$
and $W$ are covariant derivatives on ${\cal N}=2$ superspace
since they transform covariantly under the $U(2L+1|2L)$ gauge 
transformations.

Because of the constraints (\ref{constraint1}) and (\ref{constraint2}),
$\tilde{\psi}_\alpha$ is the only independent supermatrix and
the others can be solved in terms of $\tilde{\psi}_\alpha$;
\begin{eqnarray}
  && \tilde{A}_i = 
  (\sigma_i C)_{\alpha\beta}
  \left(
   2\left\{d^{(L)}_\alpha, \tilde{\psi}_\beta\right\}
   +\left\{\tilde{\psi}_\alpha, \tilde{\psi}_\beta\right\}
  \right), \\
 && \tilde{\varphi}_\alpha =
  -\left[d^{(L)}_\alpha, \tilde{W}\right]
  +\left[\gamma^{(L)}, \tilde{\psi}_\alpha\right]
  -\left[\tilde{\psi}_\alpha, \tilde{W}\right], \\
 && \label{W}
  C_{\alpha\beta}\left[d^{(L)}_\alpha, \tilde{\psi}_\beta\right]
  + \frac{1}{4}\left\{\gamma^{(L)}, \tilde{W}\right\}
  + C_{\alpha\beta}\tilde{\psi}_\alpha\tilde{\psi}_\beta
  +\frac{1}{4}\tilde{W}^2=0.
\end{eqnarray}
$\tilde{\psi}_\alpha$ is an odd supermatrix which can be expanded by
$l^{(L)}_i$ and $v^{(L)}_\alpha$, thus it is regarded as a spinor
superfield on the fuzzy supersphere.
The supermatrix $\tilde{W}$ can not be explicitly solved in general  
because of the quadratic term but we will show that it can 
be solved in a commutative limit.
For a while we treat both of $\tilde{\psi}_{\alpha}$
and $\tilde{W}$ as independent variables.
 
Though fixing the classical background as above 
violates the original $osp(1|2)$ supersymmetry (\ref{osp12}), 
it can be compensated by appropriate $u(2L+1|2L)$ gauge transformations.
Actually the action is invariant under the following combined
transformations, 
\begin{eqnarray}
 \label{susy1}
 && \delta \tilde{\psi}_\alpha 
  = \frac{1}{4}\lambda_\alpha\tilde{W}
  + C_{\beta\gamma}\lambda_\gamma
  \left\{v^{(L)}_\beta, \tilde{\psi}_\alpha\right\}, \\ 
 && \delta \tilde{W} = C_{\alpha\beta}\lambda_\beta\tilde{\psi}_\alpha
  + C_{\alpha\beta}\lambda_\beta
  \left[v^{(L)}_\alpha, \tilde{W}\right].
\end{eqnarray}
The action is also invariant under gauge transformations, 
\begin{eqnarray}
 \delta \tilde{\psi}_\alpha 
  = \left[d^{(L)}_\alpha + \tilde{\psi}_\alpha, H\right], \qquad
 \delta \tilde{W} = \left[\gamma^{(L)}+\tilde{W}, H\right],
\end{eqnarray}
for $H\in u(2L+1|2L)$.
These transformations are compatible with the constraints.
The fact that the model has the $osp(1|2)$ supersymmetry even after
choosing the classical background can be also understood as the following. 
The elements with the form 
$u_i l^{(L)}_i + C_{\alpha\beta} \lambda_\alpha v^{(L)}_\beta$
in $u(2L+1|2L)$ constitute the $osp(1|2)$ subalgebra.
Thus the model originally has two independent $osp(1|2)$ symmetries.
The classical background preserves a half of these symmetries which is given
by the transformations (\ref{susy1}). 

Mapping from the supermatrix model to a noncommutative field theory on
the supersphere is performed by the same method as ones used in \cite{IU}.
The classical solutions $l^{(L)}_i$ and $v^{(L)}_\alpha$ are mapped to
coordinates ($x_i, \theta_\alpha$) on the supersphere,
\begin{eqnarray}
 x_i &=& \frac{1}{\sqrt{L(L+1/2)}} \ l^{(L)}_i, \\
 \theta_\alpha &=& \frac{1}{\sqrt{L(L+1/2)}} \ v^{(L)}_\alpha,
\end{eqnarray}
where we set a radius of the supersphere to 1 for simplicity, 
that is $x_ix_i+C_{\alpha\beta}\theta_\alpha\theta_\beta=1$. 
And $d^{(L)}_\alpha$ and $\gamma^{(L)}$
are written as second order polynomials of $x_i$ and $\theta_\alpha$
according to (\ref{gamma}) and (\ref{d}). 
The adjoint actions of these $osp(2|2)$ generators are mapped to the
differential operators on the supersphere.
The supertrace becomes an integration on the supersphere,
\begin{equation}
{\rm STr}_L \longrightarrow -\frac{1}{2\pi}\int d^3x d^2\theta \ 
 \delta (x^2+\theta^2-1),
\end{equation}
where $\theta^2=C_{\alpha\beta}\theta_\alpha\theta_\beta$.
The supermatrix $\tilde{\psi}_\alpha$ is mapped to a spinor
superfield which  can be regarded as the spinor connection on the
fuzzy supersphere.
The field theory action derived by these procedure 
is written in terms of this spinor superfield.
The construction of the supersymmetric gauge theory on the fuzzy
supersphere given here is a natural extension of the covariant
superspace approach for the ordinary super Yang-Mills theory in 
a flat space-time.   

Next we consider a commutative limit to see that our model
is indeed noncommutative generalization of the ordinary 
supersymmetric gauge theory on sphere.
A commutative limit is given by taking $L\rightarrow\infty$ limit
keeping the radius of the supersphere fixed.
In this limit superfields $\tilde{\psi}_\alpha$ and $\tilde{W}$ are
expanded as
\begin{eqnarray}
 \tilde{\psi}_\alpha (x, \theta) &=& \eta_\alpha(x) 
  + (\sigma_\mu)_{\beta\alpha}a_\mu(x) \theta_{\beta}
  + \left(\xi_\alpha(x) 
     +\frac{1}{2r^2}x_i\partial_i\eta_\alpha(x)\right)\theta^2, \\
 \tilde{W}(x, \theta) &=& 
  w(x) + C_{\alpha\beta}\zeta_\alpha(x)\theta_\beta+
  \left(F(x)+\frac{1}{2r^2}x_i\partial_i w(x)\right)\theta^2,
\end{eqnarray}
where $r^2=x_ix_i$ and $\mu=0, 1, 2, 3$.
$a_\mu, w, F$ are bosonic and 
$\eta_\alpha, \xi_\alpha, \zeta_\alpha$ are fermionic fields.
The $u(2L+1|2L)$ gauge parameter is also expanded as 
\begin{equation}
 H(x, \theta)=h(x)+C_{\alpha\beta}h_\alpha(x)\theta_\beta+f(x)\theta^2.
\end{equation}
The gauge transformation generated by $h(x)$ is an ordinary
gauge transformation while the others are supersymmetric
extension of it.
To fix these extra gauge degrees of freedom 
generated by $h_{\alpha}$ and $f$, we set 
the Wess-Zumino like gauge fixing condition 
$C_{\alpha\beta}\theta_\alpha\tilde{\psi}_\beta=0$, thus 
$\eta_\alpha = a_0 = 0$. 
In the commutative limit the constraint (\ref{W}) can be solved as
$\tilde{W}=2x_i(\sigma_iC)_{\alpha\beta}\theta_\alpha\tilde{\psi}_\beta$, 
so that
\begin{equation}
 w=\zeta_\alpha=0, \qquad 
  F=-2(x\cdot a)\theta^2.
\end{equation}
Here we used the fact that the fields $\tilde{\psi}_\alpha$ and
$\tilde{W}$ scale as ${\cal O}(1)$ while $d^{(L)}_\alpha$ and
$\gamma^{(L)}$ scale as ${\cal O}(L)$ in $L\rightarrow\infty$.
By the gauge fixing condition and the constraint, the independent fields 
are now $a_i(x)$ and $\xi_{\alpha}(x)$.

We then decompose $a_i$ into a scalar field $\phi$ and a gauge field
$a_i^{(2)}$ on $S^2$,
\begin{equation}
 a_i = n_i \phi + a_i^{(2)},
\end{equation}
where $n_i$ is a unit normal vector on the sphere and 
$n_ia^{(2)}_i=0$ and $a_i^{(2)}$ is a tangential projection; 
$a_i^{(2)}=(\delta_{ij}-n_i n_j)a_j$.
The $U(1)$ field strength for the tangential component 
is defined as 
$F^{(2)}_{ij}=R_ia^{(2)}_j-R_ja^{(2)}_i-i\epsilon_{ijk}a^{(2)}_k$ 
where $R_i=-i\epsilon_{ijk}n_i\frac{\partial}{\partial n_i}$ are 
derivatives on $S^2$.
After straightforward calculations, we obtain the action in the
commutative limit,
\begin{eqnarray}
 \label{com-action}
 S &=& \frac{L'(L'+1/2)}{4\pi}\int d\Omega
  \left\{
   \frac{9}{8}F^{(2)}_{ij}F^{(2)}_{ij}
   -\frac{i}{4}\left(\epsilon_{ijk}n_i F^{(2)}_{jk}\right)\phi
   +\frac{9}{4}\left(R_i\phi\right)^2
   -\frac{1}{4}\phi^2 \right. \nonumber \\
 && -\frac{9}{4}(\sigma_iC)_{\alpha\beta}\xi_\alpha R_i \xi_\beta
  -\frac{1}{4}C_{\alpha\beta}\xi_\alpha\xi_\beta  \nonumber \\ 
 && \left. 
     -3 \beta 
     \left[
      i\left(\epsilon_{ijk}n_iF^{(2)}_{jk}\right)\phi
      +\phi^2
      +C_{\alpha\beta}\xi_\alpha\xi_\beta ,
     \right] 
    \right\}.
\end{eqnarray}
This model consists of a $U(1)$ gauge field which has no local degrees
of freedom, a scalar and a Majorana fermion.
There is a free tunable parameter $\beta$.
Mass of the fermion is given by $\sqrt{\mu(\mu-1)}$ with 
$\mu=(1+12\beta)/9$ because of 
\begin{equation}
 \langle \xi_\alpha(x)\xi_\beta(y)\rangle \sim
  \frac{(C\sigma_i)_{\alpha\beta}R_i+(1-\mu)C_{\alpha\beta}}
  {R_iR_i-\mu(\mu-1)}.
\end{equation} 
The massless Dirac operator is given by $D=\sigma_i R_i +1$. 
The second term in $D$ comes from the spin connection on the 
sphere and this $D$ anticommutes with the chirality operator
$\sigma_i n_i$. 
By integrating out the scalar field $\phi$ it can be shown 
that the bosonic supersymmetric parter of $\xi$ has the same mass. 
For $\beta=2/3$ the action becomes the one given in \cite{IKTW}. 
On the other hand, for $\beta=-1/12$ the mixing term between $\phi$ and
$a^{(2)}_i$ disappears.
For both cases, bosonic and fermionic excitations are 
massless.

Since the gauge fixing condition we chose is not invariant under
the supersymmetry transformations (\ref{susy1}), we must compensate them
by the field-dependent gauge transformation with the following gauge
parameter,
\begin{eqnarray}
 &&h_\alpha=-\lambda_\alpha\phi
  -i\epsilon_{ijk}(\sigma_k)_{\beta\alpha}\lambda_\beta n_i a^{(2)}_j,
  \\ 
 && f=\frac{1}{2}(\sigma_iC)_{\alpha\beta}\lambda_\beta n_i \xi_\alpha.
\end{eqnarray}
Then the supersymmetry transformations become
\begin{eqnarray}
 \delta_\lambda \phi &=& 
  -\frac{1}{2}\left(C_{\alpha\beta}\lambda_\beta\xi_\alpha\right), \\
 \delta_\lambda a^{(2)}_i &=& -\frac{i}{2}\epsilon_{ijk}n_j
  \left(\sigma_kC\right)_{\alpha\beta}\lambda_\beta\xi_\alpha, \\
 \delta_\lambda \xi_\alpha &=& 
  -\frac{1}{2}\lambda_\alpha
  \left(\phi+\frac{i}{2}\epsilon_{ijk}n_i F^{(2)}_{jk}\right)
  +\frac{1}{2}(\sigma_i)_{\beta\alpha}\lambda_\beta R_i\phi.
\end{eqnarray}
The $U(1)$ gauge transformation is given by
\begin{eqnarray}
 \delta a^{(2)}_i = R_i h, \qquad
  \delta \phi = \delta \xi_\alpha =0.
\end{eqnarray}
The action (\ref{com-action}) is invariant under these transformations.
It can be shown that the commutation relations between two supersymmetry 
transformations generate the translations on $S^2$ up to gauge
transformations, 
\begin{eqnarray}
 \left[\delta_{\lambda^1}, \delta_{\lambda^2}\right] \phi &=&
  \Theta_i R_i \phi, \\
 \left[\delta_{\lambda^1}, \delta_{\lambda^2}\right] a^{(2)}_i &=&
  \Theta_j R_j a^{(2)}_i + i\epsilon_{ijk}\Theta_j a^{(2)}_k
  +R_i\left[\Theta_j(n_j\phi- a^{(2)}_j)\right], \\
 \left[\delta_{\lambda^1}, \delta_{\lambda^2}\right] \xi_\alpha &=&
  \Theta_iR_i\xi_\alpha 
  + \frac{1}{2}\Theta_i(\sigma_i)_{\beta\alpha}\xi_\beta,
\end{eqnarray} 
with 
$\Theta_i=-\frac{1}{2}(\sigma_iC)_{\alpha\beta}
\lambda^2_\alpha\lambda^1_\beta$. 
We again note that the action has these symmetries independently of the
parameter $\beta$.

In this paper we concentrated on the case of $U(1)$ gauge theory, 
but it is easily generalized to supersymmetric gauge theory with 
$U(k) \ (k>1)$ group.

\section{Summary and discussions}
In this paper, we constructed a supersymmetric gauge theory on 
fuzzy supersphere from a supermatrix model. Our construction 
of the supermatrix model is 
based on the prescription given by Klimcik \cite{Klimcik}.
We have shown that the fuzzy supersphere can be obtained 
as a classical solution of the supermatrix model and fluctuations
around it become supersymmetric gauge fields on the supersphere.
Namely, both of the Klimcik's gauge theory and fuzzy supersphere
background are shown to be derivable from a single supermatrix.
In this sense, this model has background independence
as well as 
the other known gauge theories on noncommutative space-time.

The formulation adopted here is similar to the so called
covariant superspace approach to supersymmetric gauge theories
on superspace. Namely we first introduce larger degrees of 
freedom corresponding connection superfields on the superspace
and then impose various constraints on the superfields.
In our case, we follow the prescription by Klimcik and 
start from connection superfields 
on ${\cal N}=2$ superspace on the sphere to obtain ${\cal N}=1$
supersymmetric field theory.
In this way, we can impose appropriate constraints to eliminate
redundant fields. In our previous paper \cite{IU}, we started
from less number of superfields corresponding to 
connections on ${\cal N}=1$ but could not impose the appropriate
constraint compatible with $osp(1|2)$ global supersymmetry.
This is in contrast with the ordinary flat case where 
we can impose appropriate constraints on the connection
superfields on ${\cal N}=1$ superspace.

Our construction may be generalizable to higher dimensional
curved cases. For example, it will be interesting to 
consider a supermatrix model based on $psu(2|2)$ super Lie 
algebra to obtain a supersymmetric 
gauge theory on noncommutative superspace of $AdS^2 \times S^2$.

\vspace{5mm}
{\bf Acknowledgments}\\
We would like to thank C. Klimcik for informing us of his paper
\cite{Klimcik}. 
The work of H.U. is supported in part by JSPS Research Fellowships for
Young Scientists.


\end{document}